\documentclass[a4paper,11pt]{article}
\usepackage{pos}
\usepackage{xspace}
\usepackage{subcaption}
\usepackage{booktabs}
\usepackage{listings}

\renewcommand{\imath}{\mathrm{i}}
\newcommand{\muu}[1]{\mu_u^{\text{eff,}#1}}

\newcommand{\FS}{\texttt{Flex\-ib\-le\-SUSY}\@\xspace}
\newcommand{\FD}{\texttt{Flex\-ib\-le\-Decay}\@\xspace}

\newcommand{\SLHA}{\texttt{SLHA}\@\xspace}

\newcommand{\HT}{\texttt{Higgs\-Tools}\@\xspace}
\newcommand{\HS}{\texttt{Higgs\-Signals}\@\xspace}
\newcommand{\HB}{\texttt{Higgs\-Bounds}\@\xspace}

\title{Constraining Higgs sectors of BSM models - the case of 95 GeV "Higgs"}
\ShortTitle{95 GeV "Higgs" in the MRSSM}

\author*[a]{Wojciech Kotlarski}
\author[b]{Jan Kalinowski}

\affiliation[a]{National Centre for Nuclear Research, Pasteura 7, 02-093 Warsaw, Poland}

\affiliation[b]{
Faculty of Physics, University of Warsaw, Pasteura 5, 02-093 Warsaw, Poland
}

\emailAdd{wojciech.kotlarski@ncbj.gov.pl}
\emailAdd{jan.kalinowski@fuw.edu.pl}

\abstract{

In view of lack of the direct experimental evidence for a Beyond the Standard Model (BSM) physics, accommodating a SM-like Higgs boson is on the most important constraints that a BSM model must fulfill.
Already for some time the \texttt{FlexibleSUSY} spectrum generator generator allowed for a reliable prediction of masses and decay patterns of the BSM Higgs boson in a large class of user defined supersymmetric and non-supersymmetric models.
However, no easy way to compare those predictions with experimental data existed.
To that end we present here an interface between \texttt{FlexibleSUSY} and \texttt{HiggsTools}, a computer program assessing in a statistically meaningful way consistency of a BSM Higgs sector with experiments.

Motivated by the recent ATLAS and CMS observation of the di-photon excess at a mass of $\sim$95~GeV we demonstrate the capabilities of our framework by investigating whether the observed low mass excesses around 95 GeV seen in the data can be explained as the lightest scalar of the Minimal R-symmetric Supersymmetric Standard Model, without spoiling the SM-like properties of the second-to-lightest state.
We also briefly comment on the light dark matter candidate which is a necessarily ingredient of such a setup.
}

\FullConference{Corfu Summer Institute 2023 "School and Workshops on Elementary Particle Physics and Gravity" (CORFU2023)\\
 23 April - 6 May , and 27 August - 1 October, 2023\\
Corfu, Greece\\}

\begin{document}
\maketitle

\section{Introduction}

The discovery of a 125 GeV Higgs boson at the LHC in 2012 and the subsequent precise measurement of its properties put a strong test on any potential Beyond the Standard Model (BSM) theory.
However, to capitalize on this discovery one has to be able to predict those observables in a BSM model with an equally impressive accuracy. 
While precise computations of various observables existed for a selection of popular BSM models, the analysis of any other extension was troublesome.

To that end \FS \cite{Athron:2014yba,Athron:2017fvs} was developed to enable automatized phenomenology in a large class of user defined BSM models.
\FS capabilities include, among many other observables, precise predictions of Higgs boson masses.
Furthermore, \FD \cite{Athron:2021kve} extended \FS capabilities by allowing to compute decays of scalar particles.
The special emphasis was put on the  Higgs-like states, with the goal being to reach precision comparable with current experimental measurements.
Here we describe a recent update to \FD module.
To streamline the application of experimental constraints from Higgs searches we created an interface to \HT \cite{Bahl:2022igd}, a computer program which encodes searches for additional scalars and measurements of the detected Higgs boson at 125 GeV.
This allows to derive a statistically meaningful information on the validity of a given parameter point in relation to currently available experimental data.

We illustrate the application of our work by investigating an intriguing question of whether there are additional Higgs-boson like scalar states, possibly with mass smaller than 125 GeV.
Such scalar particles below the SM Higgs boson mass have been searched for in the past at LEP and Tevatron and now at the LHC, showing some promising deviations from the SM.

Such states occur frequently in BSM theories.
For example, it has been show that in the Minimal R-symmetric Supersymmetric model (MRSSM) \cite{Kribs:2007ac} the SM-like Higgs boson can be realised as a second-lightest scalar particle leaving room for a lighter state below 125 GeV \cite{Diessner:2015iln}.   
The MRSSM by itself is an attractive alternative to the MSSM, elegantly solving some of its shortcomings while also significantly altering its phenomenology.
The absence of left-right squark mixing eliminates many parameters responsible for flavour-violating interactions and thus alleviates the SUSY flavour problem.
Its various aspects, including Higgs physics, electroweak precision observables, flavour physics, Dirac gauginos, color octet scalars and dark matter have been studied in a long series of papers \cite{Bertuzzo:2014bwa, Diessner:2014ksa, Diessner:2015yna, Diessner:2015iln, Athron:2022isz,Belanger:2009wf, Chun:2009zx, Buckley:2013sca,Choi:2008ub,Plehn:2008ae,Goncalves-Netto:2012gvn,Kotlarski:2016zhv,Darme:2018dvz,Choi:2010gc,Choi:2009ue,Chalons:2018gez,Dudas:2013gga, Fok:2010vk,Kotlarski:2019muo,Borschensky:2024zdg}.

In what follows we focus on LEP and LHC hints of extra scalar state at around 95 GeV \cite{LEPWorkingGroupforHiggsbosonsearches:2003ing,CMS-PAS-HIG-17-013,CMS-PAS-HIG-20-002,ATLAS-CONF-2023-035}.
As the LEP excess is very broad, both LEP and LHC cases can be addressed simultaneously by a single state which we will identify with a mostly singlet-like state from the MRSSM.
\section{\texttt{HiggsTools} interface}
A convenient way of parametrizing the BSM Higgs sector is via effective couplings $\kappa$ \cite{LHCHiggsCrossSectionWorkingGroup:2013rie},
\begin{equation}
\label{eq:effc_def}
 \kappa^2_i \equiv \frac{\Gamma(H_i \to AB)_{\text{BSM}}}{\Gamma(h \to AB)_{\text{SM},m_h = m_{H_i}}}\,,
\end{equation}
defined as ratios of BSM partial widths to SM widths for the SM Higgs of the same mass.
Such couplings can be then used as input to \HT, which then internally recomputes partial widths and cross sections and quantifies the viability of a given parameter point in relation to current experimental data.

In this section we describe the construction of such effective couplings in \FS and the process of linking it with \HT.
This interface is planned to be part of 2.9.0 release of \FS (the proper paper documenting it is in preparation) while a preliminary version can be already obtained from \href{https://github.com/FlexibleSUSY/FlexibleSUSY/tree/feature-HiggsTools-interface}{github}.

\subsection{Construction of effective couplings}

For the calculation of effective couplings in Eq.~\eqref{eq:effc_def} one has to construct a hypothetical SM with the Higgs boson of mass equal to the BSM Higgs state.
\FS already contains a built-in SM in addition to  the BSM model created by the user.
For the constuction of such a SM equivalent state,  a quartic coupling paramet $\lambda$ is tuned to  provide the SM Higgs mass $m_h$ equal to a BSM Higgs mass $m_{H_i}$ (for CP-even Higgs) or $m_{A_i}$ (for CP-odd Higgs).
The case with undefined CP prorties (as happens in CP-violating models) is also correctly handled.
This procedure is done separately for every state $H_i$ or $A_i$.
Since for $m_h \gtrsim 650$ GeV the quartic coupling $\lambda$ enters a non-perturbative regime, this procedure limits the application of our interface to BSM Higgs masses in the range $[1, 650]$ GeV.
BSM states beyond that mass range are not passed to \HT.

\subsection{\texttt{HiggsTools} interface}

\HT allows for two ways to compare theory prediction with experimental data, by either specifying cross-sections and branching ratios directly or providing effective coupling description.
As calculation of cross-sections is beyond the scope of codes like spectrum generators we choose to use the effective coupling input.

The procedure of calling \HT is transparent to the user.
The user only has to point to the location of \HT during configuration of \FS using two new configure flags
\begin{lstlisting}[language=bash]    
$ ./configure --with-higgstools-incdir=... --with-higgstools-libdir=...
\end{lstlisting}
which are used to specify the location of \HT header files and library, respectively, which we mark with dots in the above code snippet.
Those should be replaced with actual paths depending on \HT installation on user's computer.

At runtime, the user has to specify the location of \HS and \HB databases using two new command line options (we use the single extended SM (\texttt{SSM}) in this example call):
\begin{lstlisting}[language=bash]    
$ ./run_SSM.x --higgsbounds-dataset=... --higgssignals-dataset=...
\end{lstlisting}
We refer to \HT documentation \cite{Bahl:2022igd} as to how to obtain and install both the library and respective databases.

The output of \HT is then shown in two new \SLHA-like \cite{Skands:2003cj,Allanach:2008qq} blocks printed out by \FS.
The new \HS block looks for example as follows 
\begin{lstlisting}
Block HIGGSSIGNALS
     1     1.59000000E+02   # number of degrees of freedom
     2     1.68975284E+02   # chi^2
     3     1.52489682E+02   # SM chi^2 for mh = 125.090000 GeV
     4     2.63146088E-04   # p-value
\end{lstlisting}
and contains in the first two lines the number of degrees of freedom from the number of observables in \HT database and the $\chi^2_{\text{BSM}}$ for the BSM model reported by \HT.
To allow for a quick interpretation of results we also provide as a reference a $\chi^2_{\text{SM}}$ for a SM and $p$-value computed for $\chi^2_{\text{BSM}} - \chi^2_{\text{SM}}$ difference assuming a $2d$ distribution.
This can be enough in many cases to determine the validity of a concrete parameter point.
If more sophisticated statistical treatment is needed, one can refer  directly to $\chi^2_{\text{BSM}}$.

For the \HB we provide the ratios $r$ to the 95\% excluded cross sections in the second new block, which may look like
\begin{lstlisting}
Block HIGGSBOUNDS
 25  1     4.03838082E-01   # LEP [eeHZ>bb] from hep-ex/0306033 (LEPComb 2.5fb-1, M=(12, 120))
 25  2     8.36015679E-01   # expRatio
 35  1     6.60215252E-01   # LHC8 [vbfH,HW,Htt,H,HZ]>[bb,tautau,WW,ZZ,gamgam] from CMS-PAS-HIG-12-045 (CMS 17.3fb-1, M=(110, 600))
 35  2     3.32131274E+00   # expRatio
\end{lstlisting}
The first column contains the PDG ID of a BSM Higgs-like state while the second distinguishes between the observed (1) and expected (2) limits.
The analysis providing the strongest exclusion is given in the \SLHA comment of entry 1.
Points with $r_{\text{obs}} > 1$ should be considered by the user as excluded at 95\% C.L..
\section{Application: The case of 95 GeV Higgs in the MRSSM}

We now briefly discus an example application of the interface described in the previous section.
During recent years, a collection of evidence hinting at a low mass, Higgs-like resonance between 95--100 GeV emerged.
These excesses can be accommodated for example in models where the SM-like Higgs mixes with an extra gauge-singlet field, as is the case in the MRSSM.

In this section we summarize the aforementioned hints and show how they can be realised in the MRSSM.
Thanks to the work described in the previous section, we prove that we still obtain a SM-like second-to-lightest Higgs boson.
We also comment on light dark matter, a necessary prediction of such a setup and general collider constraint.
This analysis is described in details in Ref.~\cite{Kalinowski:2024uxe} to which we refer the reader for more details.

\subsection{Hints for a 95 GeV Higgs}

There is a long standing LEP anomaly observed in the $Z b \bar{b}$ final state in the $95 \lesssim m_{b\bar{b}} \lesssim 100$ invariant mass window \cite{LEPWorkingGroupforHiggsbosonsearches:2003ing}.
This anomaly can be explained for example by a scalar state $s$ with a mass of 98 GeV whose combined production and branching ratio is roughly an order of magnitude smaller than that of a SM-like Higgs $h$ of the same mass \cite{Cao:2016uwt}
\begin{equation}
\mu_{Zb\bar{b}} = \frac{\sigma\left(e^+ e^- \to Z^* s \to Z b\bar{b} \right)}{\sigma\left(e^+ e^- \to Z^* h^\text{SM}_{98}\to Z b \bar{b} \right)}  = 0.117 \pm 0.057.
\end{equation}
The data collected by  CMS in Run 1 and the first year of Run 2 showed a  di-photon excess, recently confirmed  based on full Run 2 data set  together with  a small deviation seen in ATLAS \cite{CMS-PAS-HIG-17-013,CMS-PAS-HIG-20-002,ATLAS-CONF-2023-035,Biekotter:2023oen}
\begin{align}
 \mu_{\gamma \gamma}^\text{ATLAS+CMS} = \frac{\sigma(gg\to s \to \gamma \gamma)}{\sigma (gg \to h^\text{SM}_{95.4} \to \gamma \gamma)} = 0.24^{+0.09}_{-0.08}.
\end{align}

These observations  have triggered speculations that they could arise from production of a new particle (see e.g. \cite{Biekotter:2023jld,Cao:2023gkc} and references therein).

\subsection{A light CP-even scalar in the MRSSM}

In the MRSSM the CP-conserving Higgs boson is a mixture of four states: $H_u$, $H_d$, $S$ and $T$.Apart from the known MSSM-like $SU(2)_L$ doublets $H_u$, $H_d$ it features additionally an $SU(2)_L$ singlet ($S$) and a triplet ($T$).
As the triplet should be rather heavy due to the  electroweak precision observables (\cite{Diessner:2014ksa}) and SM-like Higgs state should be $H_u$ dominated, it is enough to focus here on the 2x2 mass submatrix in $H_u$ -- $S$ space (we refer to ~\cite{Kalinowski:2024uxe} for notation and details) 
\begin{align}
\mathcal{M}^{\phi}_{u,S}&=
\begin{pmatrix}
m_Z^2 +\Delta m^2_{rad}& v_u \left(\sqrt{2} \lambda_u \muu{-} +g_1 M_B^D\right) \\
v_u \left(\sqrt{2} \lambda_u\muu{-} +g_1 M_B^D\right) \; &
4(M_B^D)^2+m_S^2+\frac{\lambda_u^2 v_u^2}{2} \;\\
\end{pmatrix}
\;,
\label{eq:hu-s-matrix}
\end{align}
A light, mostly signet scalar with small mixing to the SM-like Higgs requires
\begin{align}
  M_B^D, m_S \lesssim m_Z\; 
\end{align}
The coupling of singlet to SM fermions is generated via aforementioned small mixing, controlled by the (2,1) entry of matrix in Eq.~\ref{eq:hu-s-matrix}.
The relative increase of $\mu_{\gamma \gamma}^\text{ATLAS+CMS}$ compared $\mu_{Zb\bar{b}}$ is achieved via the interplay between scaling of partial widths and total decay width of the singlet-like state with the mixing.
Care has to be also taken to ensure that the mixing does not destabilize the SM-like properties of the $H_u$ state.
This is checked with \HT.

\begin{table}[t]
\begin{center}
\begin{tabular}{l|l|l}
& BMP7 & BMP8 \\
\midrule
\midrule
$\tan\beta$ & $49.5$ & $49.8$ \\
$B_\mu$     & $176^2$ & $142^2$ \\
$\lambda_d$, $\lambda_u$ & $-0.193, -0.00658 $ & $0.161,-0.0135 $ \\
$\Lambda_d$, $\Lambda_u$ & $1.49,-1.03$ & $1.49, -0.722$ \\
%\midrule
$M_B^D$ & $45.2$ & $42.1$ \\
$m_S^2$ & $27.4^2$ & $54.1^2$ \\
$m_{R_u}^2$, $m_{R_d}^2$& $1292^2,522^2$ & $1033^2$,$788^2$ \\
$\mu_d$, $\mu_u$ & $1536,658$ & $1500,1282$ \\
$M_W^D$ & $1458$ & $1490$ \\
$M_O^D$ & \multicolumn{2}{c}{$3000$} \\
$m_T^2$, $m_O^2$ & \multicolumn{2}{c}{$3000^2,1500^2$} \\
%\midrule
$m_{Q;1,2}^2$, $m_{Q;3}^2$ & $3803^2,3900^2$ & $1465^2, 3477^2$\\
$m_{D;1,2}^2$, $m_{D;3}^2$ & $3148^2,3728^2$ & $1456^2, 1990^2$\\
$m_{U;1,2}^2$, $m_{U;3}^2$ & $1271^2,2452^2$ & $3285^2, 3967^2$\\
$m_{L;1,2}^2$, $m_{E;1,2}^2$&$1000^2$, $1000^2$ & $1680^2$, $1022^2$\\
$m_{L;3,3}^2$, $m_{E;3,3}^2$ & $1000^2$, $1000^2$& $803^2$, $185^2$\\
\midrule
$m_{H_d}$ & $-1884^2$ & $-1711^2$\\
$m_{H_u}$ & $-1063^2$ & $-1534^2$\\
$v_S$ & $-3087$ & $2004$ \\
$v_T$ & $0.35$ & $0.0142 $\\
\midrule
\midrule
$m_{h_1}$  & 95.4 & 95.4 \\
$m_{h_2}$  & 125.25 & 124.72 \\
$m_{W^\pm}$ & 80.375 & 80.371 \\
$m_{\chi_1}$  & 44.98 & 42.65 \\
$m_{\tilde{\tau}_R}$ & 1000 & 124.7 \\
$\rho^\pm_1$ & 717 &  1310 \\
$m_{a}$  & 24.85 & 54.20 \\
\bottomrule
\end{tabular}
\end{center}
\caption{
Benchmark points for the scenario discussed here: input parameters, parameters determined via tadpole equations and selected predicted, phenomenologically relevant, pole masses. Dimensionful parameters are given in GeV or GeV${}^2$, as appropriate.
Input values %,  rounded to 3 or 4 significant digits 
are listed in the upper part of the table, while derived masses of some  light physical states are in the lower part.
}
\label{tab:BMP}
\end{table}

\begin{figure}
 \centering
 \begin{subfigure}
 {0.49\textwidth}
 \includegraphics[width=\textwidth]{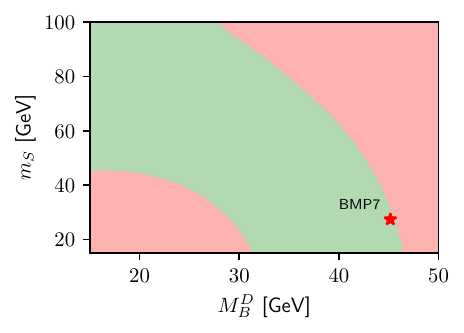}
 \caption{}
 \label{fig:hs_bmp1_M1_mS}
 \end{subfigure}
 \begin{subfigure}
 {0.49\textwidth}
 \includegraphics[width=\textwidth]{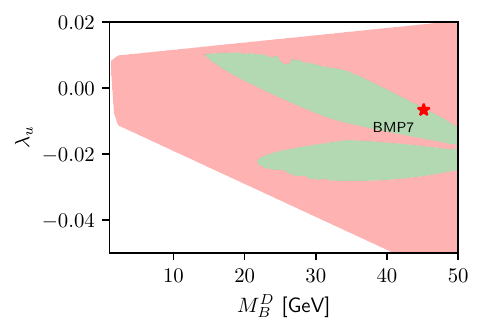}
 \caption{}
 \label{fig:hs_bmp1_M1_lamU}
 \end{subfigure}
 \caption{
    Parameter regions around BMP7 allowed (green) and excluded (red)  by SM-like Higgs data at 95\% C.L. as reported by \HS.
    White regions is where no spectrum could be generated.
 }
\label{fig:hs}
\end{figure}

\begin{figure}
 \centering
 \begin{subfigure}
 {0.49\textwidth}
 \includegraphics[width=\textwidth]{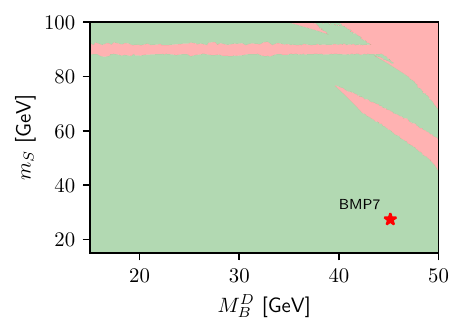}
 \caption{}
 \label{fig:hb_bmp1_M1_mS}
 \end{subfigure}
 \begin{subfigure}
 {0.49\textwidth}
 \includegraphics[width=\textwidth]{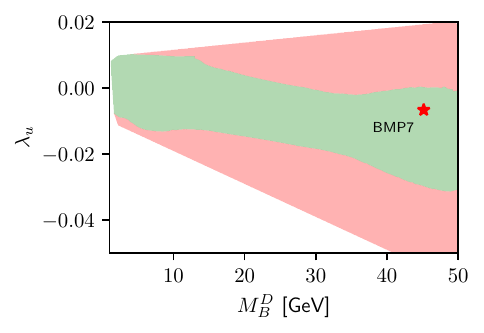}
 \caption{}
 \label{fig:hb_bmp1_M1_lamU}
 \end{subfigure}
 \caption{
    Parameter regions around BMP7 allowed (green) and excluded (red) by searches of non SM-like Higgses at 95\% C.L.  as reported by \HB.
    White regions is where no spectrum could be generated.
 }
\label{fig:hb}
\end{figure}

In Tab.~\ref{tab:BMP} we present two benchmark points consistent with  both LEP and LHC excesses simultaneously.
The points differ in the way how a proper dark matter relic density is achieved, on which we comment briefly at the end of this section.
In Fig.~\ref{fig:hs} we show $2d$ slices of the parameter space around BMP7 allowed by SM-like Higgs measurements at 95\% C.L. as reported by \HS.
Similarly, Fig.~\ref{fig:hb} shows regions excluded by searches of non SM-like Higgses at 95\% C.L. as reported by \HB.

Finally, as shown in Ref.~\cite{Kalinowski:2024uxe}, the points in Tab.~\ref{tab:BMP} are also characterized by a Bino-Singlino dark matter with a mass of around 45 GeV.
Both BMP7 and 8 give relic density of $\Omega h^2 = 0.121$ (either via the $Z$-boson resonance in the case of BMP7 or via the stau $t$-channel exchange in the case of BMP8) while remaining allowed by all dark matter direct detection experiments, including LUX-ZEPPELIN \cite{LZ:2022lsv} (as checked by \texttt{micrOMEGAs} v6 \cite{Alguero:2023zol}).
Constraints from direct production of SUSY particles at colliders were checked using \texttt{SModelS} v2.3 \cite{Altakach:2023tsd}.
\section{Summary and conclusions}

Higgs physics remains one of the main goals of the LHC as well as a target for future $e^+ e^-$ colliders.
Calculation of Higgs boson related observables and their comparison with experimental data is therefore instrumental in guiding our search for the theory of the Beyond the Standard Model physics.
In this note we described the recent improvement to \FS spectrum generator generator which allows for an easy assessment of the validity of BSM Higgs sectors by interfacing it with a popular program \HT.
We showcased it by analyzing excesses observed at LHC and LEP, interpreted as potential Higgs-like states at around 95 GeV in the context of the Minimal R-symmetric Supersymmetric Standard Model.
Both the LEP excess in the $b\bar{b}$ channel and the LHC one in the $\gamma \gamma$ channel can be simultaneously accommodated, giving a SM-like 125 GeV Higgs boson in agreement with experimental measurements (as reported by \HT).
We provide two benchmark points illustrating these features.
Those points also predict a correct dark matter relic density realised with a light bino-singlino particle (whose existance in the MRSSM is a necessary consequence of the existence of a 95 GeV Higgs-like state), which is also allowed by current direct detection experiments.
Both benchmark points are allowed by current collider limits.
While most states in BMP8 are fairly heavy (apart from light staus), BMP7 features an interesting prospect for future EW searches at the LHC thanks to the production and subsequnet decay of $\rho^- \to W^- \chi_1$ (with a 100\% branching ratio).

\section*{Acknowledgements}

We thank Henning Bahl and Sven Heinemeyer for their help regarding \HT, Peter Athron for his comments on the proper statistical interpretation of \HS results as well as Alexander Voigt for his constant work on \FS.

JK was supported by the Norwegian Financial Mechanism for years 2014--2021, under the grant No.~2019/\allowbreak34/\allowbreak H/\allowbreak ST2/\allowbreak00707.
WK was supported by the National Science Centre (Poland) grant SONATA No.~2020/\allowbreak38/\allowbreak E/\allowbreak ST2/\allowbreak00126 and, during early stages of this work, by the German Research Foundation (DFG) under grants number STO 876/2--2 and STO 876/4--2.

The authors are grateful to the Centre for Information Services and High Performance
Computing [Zentrum f\"ur Informationsdienste und Hochleistungsrechnen (ZIH)] TU Dresden for providing its facilities for high throughput calculations.

\appendix

\bibliographystyle{JHEP}
\bibliography{bibliography.bib}
\end{document}